\documentclass[journal]{IEEEtran}

\usepackage{cite}
\usepackage{comment}
\usepackage{amsmath,amssymb,amsfonts}
\usepackage{algorithmic}
\usepackage{graphicx}
\usepackage{textcomp} 
\usepackage{booktabs}
\def\BibTeX{{\rm B\kern-.05em{\sc i\kern-.025em b}\kern-.08em
    T\kern-.1667em\lower.7ex\hbox{E}\kern-.125emX}}

\usepackage{multicol}
\usepackage{xcolor}

\usepackage{hyperref}
\usepackage{fancyhdr}

\fancypagestyle{firststyle}
{
   \fancyhf{}
   
   \fancyhead[C]{\scriptsize{This is the post-print of the following article: S. Pizzi, C. Suraci, A. Iera, A. Molinaro and G. Araniti, ``A Sidelink-Aided Approach for Secure Multicast Service Delivery: from Human-Oriented Multimedia Traffic to Machine Type Communications," in IEEE Transactions on Broadcasting, doi: 10.1109/TBC.2020.2977512. \\Article has been published in final form at: \url{https://ieeexplore.ieee.org/document/9036866}. \\Copyright~\copyright~2020 IEEE.}}
}

\begin{document}

\title{A Sidelink-Aided Approach for Secure Multicast Service Delivery: from Human-Oriented Multimedia Traffic to Machine Type Communications}

\author{
\IEEEauthorblockN{
S. Pizzi\IEEEauthorrefmark{1}, C. Suraci\IEEEauthorrefmark{1}, A. Iera\IEEEauthorrefmark{2}, A. Molinaro\IEEEauthorrefmark{1}, G. Araniti\IEEEauthorrefmark{1}\\
\IEEEauthorblockA{
\IEEEauthorrefmark{1}DIIES Dept., University ``Mediterranea'' of Reggio Calabria, Italy,\\email: \emph{\{sara.pizzi$|$chiara.suraci$|$antonella.molinaro$|$araniti\}@unirc.it}\\ 
\IEEEauthorrefmark{2}DIMES Dept., University of Calabria, Italy, \\e-mail: \emph{antonio.iera@dimes.unical.it}\\
}}}

\maketitle

\begin{abstract}
To date, group-oriented communications have been mainly exploited for delivering multimedia services in human-oriented communications while, in future fifth generation (5G) cellular networks, objects will be the main target. 
Internet of Things (IoT) will undoubtedly play a key role in 5G networks, wherein massive machine-type communications (mMTC) feature a use case as crucial as challenging since cellular IoT connections are predicted to grow heavily in the next future. 
To boost capacity and energy efficiency, the 5G network can leverage device-to-device (D2D) communications which are recognized as an effective offloading technique. This is achieved thanks to the fact that, in legacy D2D communications, data are directly sent from one device to another, avoiding the crossing of the network. Obviously, the distributed nature of such a communication paradigm and the inherent broadcast nature of the wireless channel make it necessary to think how to secure the so called ``sidelink" transmissions.
This work proposes a protocol for the efficient and reliable management of multicast services in a 5G-oriented IoT scenario, in which security is a crucial requirement to be met. The proposed protocol is tailored to Narrowband IoT (NB-IoT) and makes use of D2D communications with the aim of improving network efficiency and optimizing network resource utilization. In addition, cyber security and social trustworthiness mechanisms are exploited to secure D2D communications.
\end{abstract}


\thispagestyle{firststyle}

\section{Introduction}
\label{sec:intro}

Fifth generation (5G) cellular systems will be hyper-connected networks mostly consisting of pervasive smart objects. If human communications have been the reference target in previous generations of mobile networks, they will be hugely overtaken by communications among objects in next-generation cellular networks. The 3rd Generation Partnership Project (3GPP) has specified a 5G system architecture aimed to support a wide set of use cases, typically grouped into three classes: enhanced mobile broadband (eMBB), ultra reliable and low latency communications (URLLC), and massive machine-type communications (mMTC) \cite{5G}. Focus of this work are mMTC communications, given such soaring demands for smart objects connectivity.

The number of cellular connections among objects belonging to the diversified world of the Internet of Things (IoT) is expected to reach 4 billions by 2024 \cite{2019ss}. In fact, ``smart-things networks'' are going to be exploited in the more disparate fields spanning from healthcare to agriculture, intelligent transportation system, education, industry, gaming, and so on. 
It follows that IoT networks will be composed of even-more-intelligent heterogeneous objects which, being typically resource-constrained, have the common feature of requiring a low energy consumption. This is why 3GPP standardized the Narrowband-IoT (NB-IoT) technology, optimized to guarantee longer battery life to resource-constrained cellular devices.

In the road towards 5G systems, the increase in the number of connected objects is complemented by the growth in demands for group-oriented communications.
In fact, although several IoT applications primarily involve the uplink (UL) direction, there are also many use cases, such as massive media distribution and software update, that require the same data to be sent from the network to groups of IoT devices, i.e., in the downlink (DL) direction. To date, group-oriented communications have been mainly exploited for delivering multimedia services in human-oriented communications, while, in 5G networks, objects will be the main target. Although typical traffic flows involving objects are related to sensing and automation activities, multimedia applications have a great potential in the IoT ecosystem. This is testified by the increasing interest towards the Internet of Multimedia Things (IoMT), an innovative concept according to which heterogeneous multimedia things can interact and cooperate to achieve multimedia-based applications/services \cite{iomt}. In fact, as also stated in \cite{multim-tob}, the target of multimedia services is shifting from traditional TV to streaming on connected devices, such as mobile devices and monitoring devices. 
Very recently, the feasibility of delivering less-demanding multimedia IoT applications over NB-IoT has been investigated in \cite{nbiot-iomt}.

Actually, multicast transmissions could allow to largely weigh down latency and energy consumption of the receiving IoT devices. 
The traditional way of serving multicast traffic is the Conventional Multicast Scheme (CMS), which assigns the group data rate based on the device that experiences the worst channel conditions.
As stated in \cite{multicast-tob}, guaranteeing all multicast receivers a similar performance experience is as necessary as challenging, since the instantaneous channel condition of each device in the multicast group varies independently. Despite using CMS all devices receive the same treatment, the transmission is heavily constrained by the cell-edge users. As a consequence, CMS fails to offer a high quality of experience (QoE), which is the quality focus of 5G networks \cite{qoe-tob}.
This is the reason why it is necessary to design effective methods for delivering multicast services over 5G networks.

The Single Cell Point to Multipoint (SC-PTM) architecture and procedures have been standardized to deliver multicast traffic within a NB-IoT cell.
Although NB-IoT and SC-PTM are the current standards for IoT, still several features need to be applied to further optimize the performance of IoT data delivery. In this regard, Machine-type Multicast Service (MtMS) is proposed in \cite{IEEE_Access} to define the proper architecture and transmission procedures to manage the MTC multicast traffic. Although this architecture is well-suited to IoT traffic, it does not take into account any security problems, even though their undoubted importance in the 5G ecosystem.

Among the enabling technologies of future 5G networks, device-to-device (D2D) communications stand out for the advantages they can bring in terms of latency, data rate, spectral and energy efficiency, thanks to the proximity between the communicating devices \cite{D2D}. 
As in \cite{D2D-tob}, D2D communications are often established as an underlay to cellular networks with the aim to meet the increasing demand for mobile data services, achieve high data rates, and reduce the traffic load on the base station.
On the other hand, a clear weakness in 5G-oriented D2D communications is undoubtedly the vulnerability to security attacks, during data exchanges. 
Security is one the main requirements expected for the next 5G mobile networks, mainly due to the fact that many actors will be involved in the provision of services, therefore, sensitive data will be exposed to different parties, some of which may not be trusted. Moreover, softwarization will be a key technique in the development of the 5G, hence, technologies such as Multi-Access Edge Computing (MEC), Network Function Virtualization (NFV) and Software-Defined Networking (SDN) are considered enabling for the future generation networks. Despite the undoubted benefits brought by these technologies, they cause the network to be exposed to many new security threats, which must be managed efficiently. Side-channel attacks, data breaches, distributed denial of service (DDoS) attacks, insecure interfaces and application programming interfaces (APIs), isolation failure, and malicious insiders are some of the most frequent attacks which focus on the vulnerabilities of 5G virtualization technologies. Many works in the literature deal with these problems and potential solutions \cite{mec_sec} \cite{sdn_sec} \cite{nfv_sec}. 

In IoT scenarios, where devices are often called upon to transmit sensitive data through insecure wireless channels, this flaw pushes to look for highly effective mechanisms for both data confidentiality and integrity guarantee, user authentication and authorization, and device protection.  
This problem has been tackled in our previous work \cite{wwic}, where we defined a protocol intended to secure D2D communications, in which D2D peers generate a private encryption key by using a trusted version of the Diffie-Hellman Key Exchange (DHKE) protocol. The public keys exchange is then mediated by the base station that acts as a trusted third party. The main drawback of the presented protocol is that there is no way to disclose the bad nature of the nodes before they exhibit their malicious behavior. Furthermore, the proposed protocol is not properly tailored to the IoT domain, in which the presence of resource-constrained devices poses specific challenges.

In this vein, this work proposes the \textit{Machine type Multicast Service with secure and trust D2D (MtMS-stD2D)} protocol, specifically designed for the highly reliable delivery of multicast traffic to a set of machines in IoT scenarios. It leverages D2D communications over sidelinks, coupled to a secure mechanism based on the DHKE protocol.
It is worth mentioning that sidelink is defined in \cite{3gpp-side} \textit{as the interface between UEs for sidelink communications, which also include the D2Ds, thus it is the link over which the communication between devices in proximity occurs.} 

The main contributions of this work are the following:
\begin{itemize}
\item the architecture presented in \cite{IEEE_Access} is enhanced by the inclusion of secure sidelinks, aimed at improving the performance of the multicast transmission, while ensuring protection of data transmitted in D2D;
\item the security protocol described in \cite{wwic} is strengthened by the possibility to estimate the reliability of nodes also before they perform a malicious behavior, thanks to information on the social relationships established among nodes in the network;
\item an analysis of the protocol feasibility in terms of energy consumed to implement the proposed solution on resource-constrained nodes of a NB-IoT network, such as those that populate IoT scenarios, is carried out in order to determine which kind of use case could take advantage from the MtMS-stD2D architecture.
\end{itemize}

The remainder of the paper is organized as follows. In section \ref{sec:backg}, the research background is presented. Section \ref{sec:symodel} illustrates the scenario under investigation, while section \ref{sec:proposal} describes in details all MtMS-stD2D procedures. Obtained results are shown in section \ref{sec:sim}, while conclusive remarks are given in the last section.

\section{Background}
\label{sec:backg}

In this section, we will provide the basics of NB-IoT, the cellular technology which is the reference of our work. Then, we will discuss how multicast transmissions are managed in NB-IoT. Afterwards, research works related to security are surveyed.

\subsection{The Narrowband-Internet of Things (NB-IoT) technology}

NB-IoT is a cellular technology, first defined in 3GPP Release 13, designed to meet the requirements of low-cost IoT devices located in weak-coverage signal areas. Energy saving, coverage extension, and capacity increase are among its main benefits. In fact, it can extend device battery lifetime by up to 10 years, improve coverage by 20 dBm compared to Long Term Evolution (LTE), and manage the massive capacity of IoT scenarios \cite{feltrin}. 

NB-IoT is not the only solution in licensed spectrum proposed for IoT applications, as two other alternative technologies actually exist: LTE for Machines (LTE-M), also introduced in 3GPP Release 13, and Extended Coverage GSM (EC-GSM), created from the GSM standard. NB-IoT stands out for the reduced amount of required resources. In fact, only 180 kHz of bandwidth can be used for both downlink and uplink. This choice makes NB-IoT compatible with other technologies. It can be implemented in a 200 kHz GSM carrier or, within an LTE carrier, using a 180 kHz physical resource block (PRB). 

NB-IoT can be deployed by choosing among three different operation modes: \textit{(i)} in \textit{standalone} mode, it is implemented as a dedicated carrier, using one of GSM; \textit{(ii)} in \textit{in-band} mode, it is deployed inside an LTE carrier, occupying one or more PRBs; \textit{(iii)} in \textit{guard band} mode, it can use the frequencies of the LTE carrier guard bands. As regards the in-band and guard-band modes, resources must be properly assigned to NB-IoT not to create interference with legacy LTE signals. Therefore, although NB-IoT is an independent radio access network technology, it is compatible with the previous ones, because it offers a good flexibility in the implementation. 

The physical layer of NB-IoT is characterized by innovative features, required for the management of the narrow bandwidth. In Release 13, the Frequency Division Duplexing (FDD) mode is required. Thus, in NB-IoT, uplink and downlink are frequency divided. In addition, half-duplex mode is supported by devices, that therefore can not simultaneously receive and transmit. In the downlink, NB-IoT supports Orthogonal Frequency-Division Multiple Access (OFDMA) with 15 kHz subcarrier spacing. A frame is composed of ten subframes, each lasting 1 ms, and each subframe is composed of two $0.5$ ms slots. Therefore, the downlink transmission scheme is the same as that of LTE. In the uplink direction, the Single Carrier Frequency Division Multiple Access (SC-FDMA) is used and both single-tone and multi-tone transmissions are feasible. For the single-tone transmission, it is possible to choose between a 3.75 kHz or 15 kHz subcarrier spacing. If the 3.75 kHz option is selected, slot will last 2 ms. Multi-tone transmission modes are the same as those of the LTE uplink: 15 kHz subcarrier spacing with $0.5$ ms time slot \cite{NB-IoT}. 

\subsection{Multicast support in NB-IoT}

SC-PTM has been standardized by 3GPP to manage multicast transmissions in NB-IoT networks. It extends the Multimedia Broadcast Multicast Services (MBMS) standard, from which it inherits many features, including procedures. 

The main nodes of the MBMS architecture are: the \textit{broadcast multicast-service center (BM-SC)}, which is the source of the multicast content and is responsible for the initialization of the MBMS session and for some security functions, such as the management of the authorizations for the MBMS subscribers; the \textit{MBMS-gateway (MBMS-GW)}, which is in charge of forwarding MBMS packets to the base station (BS) involved in service delivery; the \textit{multicell/multicast coordination entity (MCE)}, which has to manage admission control and radio resource allocation to the BS \cite{IEEE_Network}. 

MBMS service is subscription-based and it foresees the implementation of the following procedures: \textit{subscription}, \textit{service announcement}, \textit{joining}, \textit{MBMS notification}, \textit{session start}, \textit{data transfer}, \textit{session stop}, and \textit{leaving}. Devices subscribe to the network their interest in receiving MBMS services and the network periodically announces them the available services. A device, interested in receiving a certain service, joins the multicast group to which the service will be offered, through the joining procedure. Subscribed devices must constantly monitor the Multicast Control Channel (MCCH) for service information and listen for future service announcements.
Despite the many advantages that MBMS can bring to cellular networks \cite{mbmsb-tob}, this latter aspect is one of the most critical for IoT networks, since it can affect the Discontinuous Reception (DRX) cycle of the IoT devices and, therefore, cause a relevant energy wastage. For these reasons, SC-PTM procedures have to be modified in order to properly meet the requirements of the NB-IoT resource-constrained devices \cite{multicast_NB-IoT}.

\subsection{Securing communications}

In \cite{5G}, among the listed 5G requirements, improved security mechanisms, that can work effectively in the presence of a likely huge amount of data transmitted over cellular networks, are recommended. In the IoT landscape, devices often communicate sensitive data over the insecure wireless channel, thus, security and privacy requirements have to be satisfied to guarantee both data and device protection \cite{Sec_IoT}. Many works in the literature deal with the security problem in the next generation mobile networks, the most significant of which are collected in \cite{sec_survey}. Among the different covered topics, the authors highlight the most effective solutions proposed in the literature to overcome some challenging security issues. Solutions are grouped in access control, authentication, communication, and encryption areas, the same targeted by the protocol proposed in this paper. 

In the 5G network, D2D communications are a widely accepted technology for enhancing spectrum efficiency, improving network resource utilization, and extending battery lifetime \cite{D2D}. 
However, as also stated in \cite{survey_d2d_sec1}, the establishment of secure D2D communications is a critical point, because  of the additional problems caused by the fact that data are exchanged directly between devices in proximity. A malicious transmitter may decide to drop the data packets directed to the D2D receiver or may modify them, without the network or recipient being aware of the misbehaviour. 

Among proposals for securing D2D communication, an interesting solution is presented in \cite{SeDS}. This is the work that inspires the protocol presented in  \cite{wwic}, which satisfies many security requirements, such as non-repudiation, authentication, authorization, confidentiality, and integrity. It uses some security mechanisms, such as encryption, HMAC, and signature to manage the messages exchanged between the two peers involved in direct communication. To this aim, the encryption of transmitted data is performed through a symmetric (i.e., private-key) encryption algorithm. The private key is generated through an enhanced version of the DHKE protocol. The enhancement consists in charging a trusted third party (i.e., the BS) to manage the public keys necessary for the generation of the secret key. In detail, the DHKE algorithm states that each of the two peers, involved in the generation of the secret key, produces a preliminary public key to be sent to the other, so that it can calculate the same private key. According to DHKE, the peer \textit{i} can compute the secret key $K_{ij}$ as follows: 

\begin{equation} 
\label{eq:dhke1}
K_{ij}=Y_j^{Xi} \bmod q 
= (\alpha^{X_j})^{X_i} \bmod q
=\alpha^{X_i X_j} \bmod q
\end{equation}

where, $Y_j=\alpha^{X_j} \bmod q$ is the public key of peer \textit{j}, $\alpha$ is a fixed primitive element of $GF(q)$ and \textit{q} is a prime number (both known to the two involved peers), $X_i$ and $X_j$ are independent random numbers respectively chosen and kept secret by peer \textit{i} and peer \textit{j}. Similarly, peer \textit{j} computes the same secret key thanks to the knowledge of $Y_i$, that is the public key of peer \textit{i}:

\begin{equation} 
\label{eq:DHKE2}
K_{ij}=Y_i^{X_j} \bmod q
= (\alpha^{X_i})^{X_j} \bmod q
=\alpha^{X_i X_j} \bmod q
\end{equation}

In legacy DHKE, the peers directly exchange their public keys. Differently, in the proposed enhanced version, each peer sends its public key to the BS, which will forward it to the other peer, following a request coming from it and only after having verified its identity and legitimacy to request for that information. The literature on the subject confirms that these design choices are well suited to IoT scenarios, like the one examined in this paper. In fact, many research works assume the intervention of a trusted third party in the key generation mechanism between resource-constrained devices \cite{ttp} or of a centralized security framework to detect incoming attacks \cite{detector}. 
In \cite{gt-qosec}, the authors propose GT-QoSec, a game-theoretic joint optimization of QoS and security in Heterogeneous Networks (HetNet) that can also serve a large number of devices requesting various types of applications. In this scenario, the eNodeB (eNB) implements intrusion detection techniques to monitor threat levels of the network and plays a key role in ensuring adequate security ranks to each device. Regarding the choice of symmetric encryption, many works in the literature confirm that this is the best choice for resource-constrained devices. Works \cite{smart_home} and \cite{smart_toys} proposed symmetric encryption algorithms, that are less demanding in terms of energy consumption compared to an asymmetric approach.

So far, the advantages of the algorithm presented in \cite{wwic} have been illustrated in order to legitimize whether this algorithm is partly used also in the protocol presented in this work.
In fact, similarly to \cite{wwic}, we exploit an enhanced version of the DHKE protocol, where the BS acts as a trusted third party in order to avoid the man-in-the-middle attack, a well-known vulnerability of DHKE. 
In our proposal, the BS contributes to the establishment of secure D2D communications and to the detection of any malicious behaviour of devices, that may have been intentionally deployed for breaching network security. 
Since both DHKE and D2D communications are distributed by nature, the BS (already in charge of delivering data from/to devices) also exploits, at a global level, any security information gathered by devices in local data exchanges.  
In order to enhance the protocol presented in \cite{wwic}, it was necessary to think about how to reduce the number of undetected malicious devices selectable as D2D transmitters (or relay nodes). So, in addition to \cite{wwic}, this work foresees that the BS implements a careful selection of the D2D transmitters by considering, not only the malicious behavior of the nodes that have already played the role of relays, but also taking into account the ``social'' reputation of the devices within the network.

\section{System Model}
\label{sec:symodel}

This work considers an IoT scenario, wherein the MTC multicast traffic is managed through the proposed algorithm, named \textit{MtMS with secure and trust D2D (MtMS-stD2D)}. D2D communications are established between devices directly served by the BS and those terminals excluded from the multicast transmission, because of their adverse channel conditions. A secure protocol is implemented over sidelinks in order to protect the transmitted data. The protocol aims to avoid giving a forwarding role to devices that exhibited a malicious behavior in the past. Since a node cannot be considered as ``not secure'' (i.e. unreliable) until it behaves maliciously, the protocol also estimates the reliability of network nodes by leveraging a simple, yet effective, trust model available from the literature and based on the Social IoT (SIoT) paradigm \cite{siot}.
Social relationships that can be established among nodes are: parental object relationship (POR), among objects created by the same producer in the same period; co-location object relationship (C-LOR), that affects smart things that always work in the same place; co-work object relationship (C-WOR), between objects that collaborate to achieve a common goal; ownership object relationship (OOR), that binds objects owned by the same holder; social object relationship (SOR), due to the meeting, sporadic or continuous, of the owners of the objects, that consequently get in touch.
Examples of applicative use-cases that could benefit from the presented protocol are massive media distribution, software update of a group of machines owned by a customer/tenant, or delivery of alerting messages. 

A common trend in cellular technology is to deploy femtocells which are small, inexpensive, and low-power base stations, that represent a cost-effective means of data traffic offloading from the macrocell. 
Femtocells are generally consumer-deployed and connected to their own wired backhaul connection. 
It is expected that 70\% of wide-area IoT devices will use cellular technology in 2022 \cite{ss}. Thus, it appears clear that femtocells will play a significant role in the next-to-come scenario, and will drive the fast realization of the IoT, because of their ability to provide high data rate services in a less expensive manner \cite{femto}. For this reason, we consider a femtocell, in which an \textit{home-evolved NodeB (HeNB)} provides connectivity to a small-cell of devices, thus guaranteeing latency and energy consumption reductions and improving coverage and reliability compared to the traditional macrocell. 
In particular, NB-IoT is exploited for radio links between the HeNB and devices, whereas proximity-based transmissions (i.e., D2D) are established among devices in mutual proximity. The idea we want to investigate is to offload the portion of traffic that cannot be handled by NB-IoT via short-range sidelinks. The motivation behind this choice is that we want to utilize a low-power technology for reducing energy consumption of the devices, even if we must obviate to the lack of support of D2D communication in NB-IoT. Thus, we assume that relay nodes are equipped with two radios: a NB-IoT interface, connecting the relay node to the HeNB, and an LTE-A radio, for the direct communication with cell-edge users. This assumption is realistic since IoT devices are currently equipped with a wide range of radio technologies, that include both long- and short-range connectivity, such as Long Range (LoRa), LTE, NB-IoT, Bluetooth, and LTE Cat-M1. This requirement can be seen as a further ``hardware constraint'' in the relay node selection process.

\begin{figure}[tp]
\centerline{\includegraphics[width=0.46\textwidth]{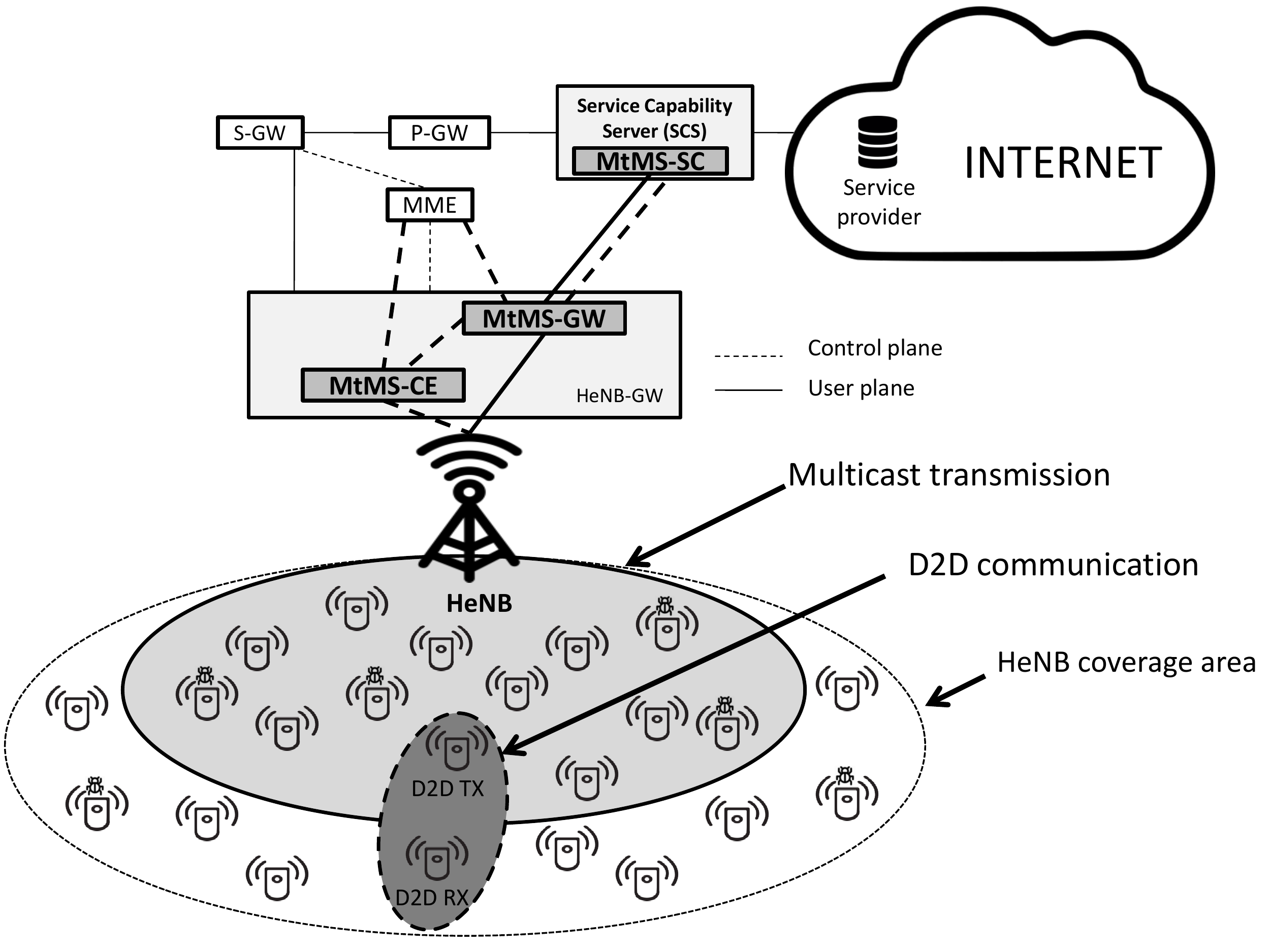}}
\caption{Reference architecture.}
\label{fig:architecture}
\end{figure}

Our MtMS-stD2D architecture, depicted in Fig. \ref{fig:architecture}, derives from the MtMS architecture (defined in \cite{IEEE_Access}) and properly enhances it to support secure D2D communications. It is composed of the following nodes: the \textit{HeNB}, which provides connectivity to a small-cell of devices; \textit{HeNB gateway (HeNB-GW)}, which aggregates control and data traffic of various HeNBs; \textit{MtMS serving center (MtMS-SC)}, implemented at the service capability server (SCS), it is responsible for initializing the MtMS session, obtaining the multicast content and the information about the receiving devices; \textit{MtMS coordination entity (MtMS-CE)}, which manages the joining procedure by paging the indicated devices; \textit{MtMS gateway (MtMS-GW)}, which receives data from the MtMS-SC and forwards them to the cells with paged devices. MtMS-GW and MtMS-CE are implemented at the HeNB-GW. 

Despite the proposed MtMS-stD2D architecture strictly relies on the presence of the HeNB, it does not pose severe scalability problems. First of all, the number of devices under coverage of the HeNB is limited, since we are considering femtocells. In addition, not all devices in the cell must implement the security protocol. In fact, only D2D communications, that involve a limited portion of devices over the total number of nodes in the femtocell, are designed to be secured. 

In view of the goal of serving resource- and energy-constrained IoT devices, we put a keen attention to the overhead introduced by the proposed security mechanism, which is then evaluated through an energy consumption analysis.

\section{Machine-type Multicast Service with secure and trust D2D}
\label{sec:proposal}

In our reference IoT environment, resource-constrained devices have to receive multicast data from the network. In many IoT scenarios, the UL direction is the most analyzed, since IoT devices are assumed to have the task of sending data to the network (sensing). Actually, use cases such as massive media distribution, software update, and delivery of alerting messages are equally important and frequent in IoT. For this reason, this paper focuses on procedures for \textit{efficient} and \textit{secure} DL service delivery. 
 
The procedures and sub-procedures of the MtMS-stD2D protocol are described below and summarized in Table \ref{tab:phases}.

\begin{table}[htb]
\centering
\caption{Procedures and sub-procedures.}
\begin{center}
\begin{tabular}{ll}
\toprule
Procedures & Sub-procedures \\
\midrule
A) Subscription		&   \\ \\

B) Initialization		&   \\ \\

C) Joining		& 1) Paging \\ 
                & 2) Random access \\
                & 3) D2D pairs selection \\
                & 4) Service request	  \\ 
                & 5) D2D pair announcement  \\ \\

D) Data transfer		& 1) Multicast transmission \\
                        & 2) stD2D communication \\
                        & 3) Report \\
                        & 4) Public key exchange \\
                        & 5) Alarm beacon or null \\ \\

E) Session stop		&  \\
\bottomrule
\end{tabular}
\end{center}
\label{tab:phases}
\end{table}

\subsection{Subscription}

The subscription procedure is performed by the service provider. For example, in the case of massive media distribution, it is realistic to assume that the owner of devices communicates to the network which devices must receive the multimedia data. This allows energy saving, since it prevents devices from interrupting their DRX cycle to monitor service announcements. 

\subsection{Initialization}

MtMS-SC is responsible for the initialization of the MtMS session. It receives data from the service provider and forwards them to the MtMS-GW. 

\subsection{Joining}

\subsubsection{Paging}

The joining procedure starts with paging, aimed at waking up the subscriber devices before data transmission. 
The MtMS-stD2D protocol includes an enhanced DRX-based group paging, which consists in the creation of subgroups of devices to be paged on the basis of their DRX cycle, at different times. The HeNB is in charge of performing paging to wake up target subgroups of devices, when necessary, since it is assumed that the network is already informed on which devices should receive the multicast data. This assumption is realistic in IoT scenarios, since one of the main applications of group-oriented communications are massive media distribution and software update. In these cases, the service provider (e.g., the owner of the devices) can communicate to the network which devices must receive the service.
The enhanced DRX-based group paging allows to: \textit{(i)} minimize the number of multicast transmissions required to deliver data to all devices in the cell, and \textit{(ii)} prevent the waste of energy caused by the interruption of the DRX cycle for service announcements monitoring.

\subsubsection{Random Access}

Whenever a subgroup is paged, the awakened devices must perform the random access procedure to synchronize with the network.
On this occasion, each device also sends information about: \textit{(i)} the conditions of the direct channel towards the HeNB (i.e., channel quality indicator (CQI)), \textit{(ii)} the CQIs of the sidelinks which connect it to the nearby nodes in the network (i.e., D2D CQIs), and \textit{(iii)} the values of social relationships with other network nodes to allow HeNB estimating nodes' reliability. 

\subsubsection{D2D pairs selection} 
\label{pairs}

This phase represents the heart of our proposal, since it involves the use of social trustworthiness and security metrics in the selection of D2D transmitters.
Based on their CQI, the network determines which devices to serve directly and which to serve in D2D. 
In selecting the best D2D transmitters for the given receivers, the network, particularly the MtMS-CE, considers both the reliability of the possible relays and the D2D CQIs. In \cite{sfn-tob}, the authors demonstrate the convenience of considering the conditions of the D2D channel in the selection process of the best transmitter for a given receiver.  
First, the MtMS-CE computes the reliability value for each possible relay as: 

\begin{equation} \label{eq:nrv}
NRV_k^t = \begin{cases} 
SRF_k^{t},  & \qquad MDC_k^{t-1}=0  \\
MDC_k^{t-1},  & \qquad MDC_k^{t-1} \ne 0  \\
\end{cases}
\end{equation}

where:
\begin{itemize}
\item $NRV_k^t$ is the Node's Reliability Value referred to node \textit{k} at time instant \textit{t}.
\item $SRF_k^t$ is the Social Relationships Factor for node \textit{k} at time instant \textit{t}. It derives from the social relationships between node $k$ and the other nodes of the network. In fact, as in \cite{fi}, a value $\in [0,1]$ is assigned to each type of the five social relationships discussed in section \ref{sec:symodel}. The kind of social relationship established between network nodes affects their reputation as long as a node exhibits a malicious behaviour when working as a D2D transmitter. For this reason, the SRF value has a great impact on the protocol performance especially in the early stages, when very few D2D communications have been set up.
\item $MDC_k^{t-1}$ is the Malicious D2D-transmissions Counter of node \textit{k} before the current time instant. More details on this counter will be given below. Briefly, it may be a value $\in [0,\infty)$, representing a measure of non-reliability of the node, since it tracks the number of times the node, selected as a D2D transmitter, behaved maliciously. 
\end{itemize}

The MtMS-CE bans as non-eligible for the role of relays all nodes for which $NRV > 1$, because they evidently behaved maliciously at least in one D2D transmission. Afterwards, it splits eligible devices in three priority classes based on their NRVs: high, medium, and low. The MtMS-CE checks the D2D CQIs between the possible relays, belonging to the high class (i.e., the most reliable), and the D2D receivers. If there is a relay for each receiver, the selection ends. Otherwise, MtMS-CE considers the medium priority class. Only if no relay is found among the nodes with medium reliability values, then MtMS-CE considers the low priority class. In case MtMS-CE cannot find a transmitter for each recipient, D2D communications are not established and all network nodes are served via CMS.  

\subsubsection{Service request}

In order to guarantee confidentiality and integrity of data transmitted over sidelinks, a secret key is generated by each transmitter and receiver by performing the DHKE protocol, as in \cite{SeDS}. The exchange, between the peers, of the public keys, required for the generation of the same secret key, is always mediated by a trusted third party, that is the HeNB. The D2D receiver, $DEV_i$, sends a service request message to the HeNB to communicate its identity and the public key generated for the implementation of the DHKE algorithm. In this and in the following messages exchanged between a device and the HeNB, the use of message authentication (i.e., HMAC) is envisioned for the integrity and authentication of each message. 

\subsubsection{D2D pair announcement}

After receiving the service request message, the HeNB authenticates the requesting device in the normal cellular communication mode, checking if its ID is registered. In the positive case, the HeNB has to perform the D2D pair announcement, informing both the D2D receiver (i.e., $DEV_i$) and the D2D transmitter (i.e., $DEV_j$) of their imminent communication. Thus, it sends to each peer the identity of the other. Furthermore, it sends to $DEV_j$ the public key received by $DEV_i$ during the service request step. 

\subsection{Data transfer}

\subsubsection{Multicast transmission}

As previously mentioned, during the initialization procedure, the service provider sends multicast data to the MtMS-SC, which forwards them to the MtMS-GW. The latter signs data before sending them to the HeNB. This way, it will always be possible to recognize the original data from the service provider. When data arrive at the HeNB, it performs the multicast transmission to the first paged subgroup by using CMS.

\subsubsection{Secure-D2D (sD2D) communication}
\label{sec:sD2D}

When $DEV_j$, which belongs to the served subgroup, receives data from the HeNB, it already knows it has to forward them to the previously notified D2D receiver; so it carries out all the operations necessary to guarantee a secure D2D communication. In order to ensure confidentiality and integrity to data packets sent over sidelink, it encrypts them with a symmetric encryption algorithm using the secret key generated through the DHKE algorithm. In addition, it reports in the message its identity (i.e., $ID_j$) and signs it before sending it to the D2D receiver. This guarantees non-repudiation.

\subsubsection{Report}

As mentioned, the HeNB acts as a trusted third party in the public keys exchange, required by the DHKE algorithm for the two peers to generate the same secret key. To this aim, the relay node sends the HeNB the public key used in the previous step to derive the encryption key.

\subsubsection{Public key exchange}

After receiving data, $DEV_i$ first verifies the identity of the transmitter. To this aim, it compares the identity reported in the message received over sidelink by $DEV_j$ (i.e., $ID_j$) with that communicated by the HeNB. If they do not match, the packet is dropped; otherwise, it proceeds with next steps. Afterwards, it checks the signature of the transmitter and, if it is valid, data are considered good because sent by the entity corresponding to $ID_j$. Once the identity of the sender is verified, $DEV_i$ needs to generate the decryption key to obtain the plaintext. Thus, it sends a public key request message to HeNB.
 
\subsubsection{Alarm beacon} 
\label{subsubsec:security}

Thanks to the reception of the public key, $DEV_i$ can get the private key and obtain the plaintext data. To verify the origin of data, it also checks the signature of the MtMS-GW and, if it is valid, data are accepted. Otherwise, it is possible that data have been tampered. In this case, $DEV_i$ must send to the HeNB an alarm beacon as the evidence of the fake message and to track the malicious attacker. The HeNB waits for the beacon for a $\Delta T$ time after sending the public key to $DEV_i$. If any alarm beacon arrives during this time interval, then the HeNB first checks the validity of the signature of the MtMS-GW. If the signature is invalid, it assumes that the message did not come from the service provider and may be fabricated by the transmitter. So, HeNB also verifies the validity of the signature of the relay node to ensure that the fake message comes from the entity corresponding to $ID_j$. One counter is stored by the HeNB to track any malicious behavior of D2D relay nodes: the MDC. In case of a malicious transmission by $DEV_j$, HeNB increments its MDC by one. This counter contributes to define the node's reliability, as explained in section \ref{pairs}.

\subsection{Session stop}

The MtMS session is completed when data are sent to all devices initially subscribed by the service provider.

\section{Performance Evaluation}
\label{sec:sim}

\subsection{Security analysis}
The features of a D2D communication make it willing to various security threats. 3GPP has released TS 33.303 in which it describes the security procedures that can be implemented in Proximity-based Services (ProSe), in particular for the public safety use case \cite{prose}. Our work aims at optimizing the encryption key generation procedures also making them eavesdroppers proof. 
Hence, we first list the security requirements for reliable D2D communications and, then, we highlight which of these requirements are met by our MtMS-stD2D protocol.
\begin{itemize}
\item \textit{Data confidentiality and integrity.} Confidentiality prevents data from being accessed by unauthorized entities. Data integrity avoids an attacker from tampering data transmitted in a private communication. 
\item \textit{Authentication.} It is important for identifying the entities that perform actions. It allows the association of identifying credentials with the entity that owns them. 
\item \textit{Privacy.} The protection of information relating to network users concerns the privacy assurance. In the age of General Data Protection Regulation (GDPR), this is a fundamental requirement in some contexts, such as eHealth and smart wearables.
\item \textit{Non-repudiation.} The non-repudiation of an action is important for the detection of malicious entities. If this requirement is met, a malicious user cannot deny having done a bad deed, so network can possibly punish it.
\end{itemize}

The main contribution to security offered by the MtMS-stD2D protocol is the \textit{reliability assurance}: thanks to the proposed selection mechanism, it is very likely that a reliable and efficient relay will be selected to forward data through D2D communications over sidelinks. As regards the protection of data packets transmitted in D2D, their \textit{confidentiality} is guaranteed by the implemented symmetric encryption algorithm, while their \textit{integrity} is assured by the use of the HMAC. The construction of the HMAC, in fact, implies that only who knows the used private key (i.e., in our protocol, the owner device and the HeNB) can modify the message; in this way, also the \textit{authentication} of the message is achieved, because only the owner of the private key can have generated it. 
Furthermore, thanks to the signature implementation, \textit{non-repudiation} is accomplished. This is particularly important in the sD2D communication step (see subsection \ref{sec:sD2D}), when the relay has to sign data before transmitting it to the D2D receiver. Once the signature has been checked, if it is successful, the relay can not deny that it was the origin of data transmission and, if it has been malicious, the network is able to take into account its bad deed. 
Finally, it is worth mentioning that, in our protocol, the exploitation of the BS as a \textit{trusted third party} is an additional security guarantee, since, by centralizing the security control, it is possible to avoid attacks by distributed devices, such as man-in-the-middle and byzantine generals problem. The fact that the BS is fundamental for the existence of the network implies that it is also the most secure and protected node, lowering the risk of central entity vulnerabilities.

\subsection{Simulation results}

We tested the performance of the proposed protocol via the Matlab tool. 

The considered scenario consists of 1000 devices distributed in the edge of a circular NB-IoT cell with a 1000 m radius. Inside the multicast group, including all terminals, a portion of devices is served according to a CMS approach through NB-IoT, while those in worst channel conditions receive data via D2D connections. 

According to NB-IoT specifications, a bandwidth of 180 kHz is available for the communication between HeNB and device, and the in-band mode is deployed. 
As regards D2D communications on LTE-A, a bandwidth of 20 MHz, which corresponds to 100 RBs, is available. A TDD LTE frame type 2 configuration 3 is used. Each slot (or Transmission Time Interval, TTI)  in the frame lasts 1 ms, so the entire frame has a duration of 10 ms. The Inband D2D mode is chosen, so uplink slots are reserved to D2D communications. In downlink slots, the multicast transmission takes place. 

Simulations are conducted by varying the percentage of malicious devices and the dimension of the downloaded file. In particular, file dimension varies from 5 kb to 10 MB in order to analyze the performance of the proposed protocol (shown in the graphs as stD2D) in different use cases, spanning from alert messaging to high dimension file downloading. Security messages dimension is set according to \cite{SeDS}. Finally, for purely simulation purposes, we consider the almost ideal situation in which devices send to the HeNB accurate values of trustworthiness, computed as in \cite{siot1}. In particular, values in the range [0,0.4] and [0,1] are respectively assigned to malicious and non-malicious nodes. The ultimate goal is to select the most reliable nodes to work as relays.

The following metrics are used to assess the performance of the proposed protocol:
\begin{itemize}
\item \textit{Percentage of wasted capacity} on the sidelink, caused by the selection of unreliable transmitters.
\item \textit{Mean number of non-corrupted received kbits}, which indicates the amount of data correctly downloaded in D2D, as transmitted by non-malicious relays.
\item \textit{Average wasted energy} by D2D receivers. In the D2D protocol (without security) the waste is caused by the reception of data sent by malicious relays; in the proposed protocol (stD2D) the waste is caused both by sending the data necessary to secure D2D communications and by receiving data sent by malicious relays.
\item \textit{Percentage of energy consumed to secure D2D communications}, computed with respect to the total energy required for the operation of D2D relays and receivers in the proposed protocol.
\item \textit{Energy consumed to download data}, computed for the stD2D protocol and in the case where data are sent directly by the HeNB to the node, in unicast mode (without D2D and security).
\end{itemize}

All these metrics are related to security assurance, since they measure at which extent the protocol is able to select reliable relays and limit the resources wastage of the devices. 

\begin{figure}[htbp]
\centerline{\includegraphics[width=0.4\textwidth]{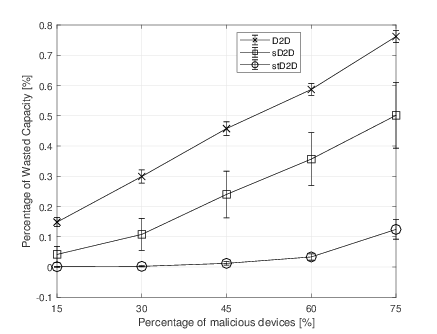}}
\caption{\% of wasted capacity vs. \% of malicious devices.}
\label{fig:data_loss_perc}
\end{figure}

\begin{figure}[htbp]
\centerline{\includegraphics[width=0.4\textwidth]{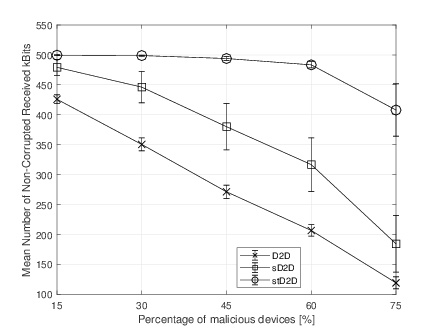}}
\caption{Amount of data correctly transmitted in D2D (i.e., by non-malicious relays) vs. \% of malicious devices.}
\label{fig:bytesUL}
\end{figure}

In Fig. \ref{fig:data_loss_perc} and \ref{fig:bytesUL} we compare three possible implementations of the D2D communication: in the \textit{D2D} case, the communication takes place without security; in the \textit{sD2D} case, the reliability of network nodes keeps into account only the respective stored MDCs; in the \textit{stD2D} case, the D2D communication is secure and the node's reliability is based both on security and social trustworthiness. We set the dimension of the file to download to 500 kb. 

Fig. \ref{fig:data_loss_perc} shows that considering both security and social trustworthiness to evaluate the reliability of devices is the winning strategy, as it guarantees practically no data loss. Fig. \ref{fig:bytesUL} confirms this claim. In fact, it shows that the proposed protocol allows to download practically the whole file of dimension 500 kb even with 60\% of malicious devices. This happens because almost only non-malicious relays are selected as forwarding nodes towards D2D receivers. 

\begin{figure}[htbp]
\centerline{\includegraphics[width=0.4\textwidth]{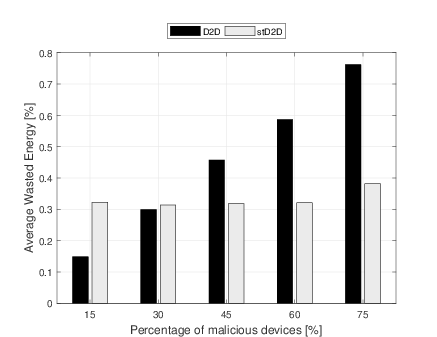}}
\caption{Avg. wasted energy by D2D receivers vs. \% of malicious devices.}
\label{fig:energy_usage}
\end{figure}

Fig. \ref{fig:energy_usage} depicts the amount of energy wasted, on average, by D2D receivers for operations other than receiving useful data. The power consumption values are set as in \cite{energy_model}. In particular, the energy in the D2D protocol (which does not provide security in direct communication between devices) is wasted by the receiver that has to download useless data, as transmitted by a malicious relay. It is computed as:

\begin{equation} \label{eq:en_wasted1}
E_{\textnormal{wasted, D2D}}=
\frac{E_{\textnormal{malicious}}}{E_{\textnormal{total}}}.
\end{equation}

As regards the stD2D protocol, the energy waste is caused by both the reception of useless data sent by a malicious relay, and the transmission of the data needed to implement the security mechanism to both HeNB and forwarding node. The energy wasted in this case is calculated as: 

\begin{equation} \label{eq:en_wasted2}
E_{\textnormal{wasted, stD2D}}=
\frac{E_{\textnormal{malicious}} + E_{\textnormal{security}}}{E_{\textnormal{total}}}.
\end{equation}

As shown in Fig. \ref{fig:energy_usage}, the energy waste for D2D protocol increases with the percentage of malicious devices in the network, as the number of D2D malicious transmitters is higher. Differently, stD2D exhibits a constant trend, indicating that the energy waste is caused mostly by the contribution $E_{\textnormal{security}}$. In fact, unlike $E_{\textnormal{malicious}}$, this does not depend on the percentage of malicious devices. 

\begin{figure}[htbp]
\centerline{\includegraphics[width=0.4\textwidth]{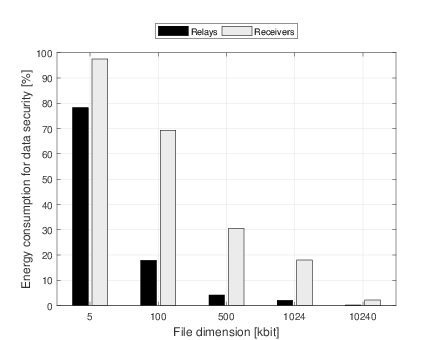}}
\caption{\% of energy consumed for data security vs. file dimension.}
\label{fig:perc_energy}
\end{figure}

Fig. \ref{fig:perc_energy} shows the total energy consumed by devices to secure D2D communications in the stD2D protocol. Results show, as expected, that securing D2D communications is a high energy-consuming task. However, it is important to point out that, without any security mechanism, some nodes will likely discharge their battery in the reception of corrupted (thus non-useful) data. This can be inferred by Fig. \ref{fig:energy_usage}. 
D2D receivers must handle more security data with respect to relays because our proposed protocol mainly requires the receivers to exchange information with the HeNB. Hence, the total energy consumption is greater for D2D receivers than for relays.

\begin{figure}[htbp]
\centerline{\includegraphics[width=0.4\textwidth]{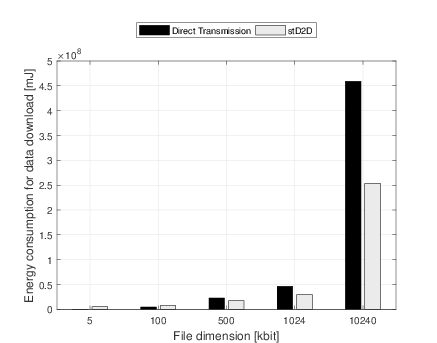}}
\caption{Energy used to download data under increasing file dimension.}
\label{fig:energy_to_down}
\end{figure}

Fig. \ref{fig:energy_to_down} shows the energy consumption computed for stD2D and in the case of direct transmission (e.g., unicast), where data are sent directly by the HeNB to cell-edge devices (i.e., D2D links are not established). The graph shows that the stD2D protocol can offer greater advantages in terms of energy saving as the file dimension increases. 

By summarizing, Fig.s \ref{fig:energy_usage}, \ref{fig:perc_energy} and \ref{fig:energy_to_down} highlight that stD2D protocol is able to guarantee the establishment of secure D2D communications while being not energy demanding, especially in case of large files. 

It is worth noting that reliability, considered as the result of both social trustworthiness and security, is the main parameter that the proposed MtMS-stD2D protocol considers in the selection process of relay nodes. Despite our approach shows the best performance results, there is the possibility that relay nodes considered unreliable are not malicious, as well as the opposite case. An ineffective relay selection can worsen the performance of the D2D transmission, as efficient relays could be discarded, due to their low reputation, leading to possible throughput wastage, or unreliable nodes could be selected as data forwarders with a consequent data loss. We can affirm that the proposed protocol is 100\% effective in the detection of malicious relays when the NRV is based on the MDC value, as only the nodes that have shown malicious behaviour in the past are considered ineligible as D2D transmitters. Differently, when NRV is based on the SRF, an error is possible in the evaluation of the nature of the nodes, since the social relationships, which determine their reputation, could provide incorrect information.

\section{Conclusions}
\label{sec:conclusions}

The Machine-Type Multicast Service with secure and trust D2D (MtMS-stD2D) protocol, proposed in this paper, manages the delivery of multicast data to a group of IoT devices. Secure D2D communications to the edge-devices are established, over sidelinks, to improve the performance of the multicast transmission for the entire network.
The security of D2D communications is guaranteed by using various means. First, D2D relays are selected based on their reliability, measured taking into account both security and social trustworthiness factors. Second, the Diffie-Hellman Key Exchange (DHKE) protocol is used to generate the secret key used for encryption and decryption of data transmitted in D2D. This guarantees data confidentiality and integrity. Third, the protocol includes an exchange of messages that allows tracking any misbehaviour by malicious D2D transmitters. 

Obtained simulation results demonstrate the effectiveness of the protocol in making a better selection of D2D transmitters, which allows to reduce data loss, thus guaranteeing almost no waste of capacity and resources. Furthermore, MtMS-stD2D proves to be not high demanding in terms of energy consumption, especially when nodes have to download large files, since it avoids that D2D receivers waste energy because of the establishment of non-secure D2D communications, thus representing an energy-efficient solution with respect to the direct transmission from the HeNB. 

Since the introduction of a trustworthiness model showed to be a significant complement to security, because of the meaningful performance improvement it can bring, future research work will be tailored to the definition of a more accurate and realistic model for evaluating nodes' reliability.

\end{document}